# Detection of Bio-aerosols and COVID-19 Equivalent Particles Via On-chip Mid Infrared Photonic Spectroscopy

ROBIN SINGH,[1,2] PETER SU[3,4], LIONEL KIMERLING[3,4,5], ANU AGARWAL[3,4,5] AND BRIAN W ANTHONY[1,2]

[1]Department of Mechanical Engineering, [2]Institute for Medical Engineering and Science,

[3]Department of Materials Science and Engineering,

[4]Microphotonics Center, [5]Materials Research Laboratory, Massachusetts Institute of Technology, Cambridge, MA, 02139, USA

We propose an on-chip mid-infrared (MIR) photonic spectroscopy platform for aerosol characterization to obtain highly discriminatory information on the chemistry of aerosol particles. Sensing of aerosols is crucial for various environmental, climactic, and pulmonary healthcare applications. Further, there are a number of unintended situations for potential exposure to bioaerosols such as viruses, bacteria, and fungi. For instance, the current pandemic scenario of COVID-19 occurring across the world. Currently, chemical characterization of aerosols is performed using FTIR spectroscopy yielding chemical fingerprinting because most of the vibrational and rotational transitions of chemical molecules fall in the IR range; and Raman spectroscopy. Both techniques use free space bench-top geometries. Here, we propose miniaturized on-chip MIR photonics-based aerosol spectroscopy consisting of a broadband spiral-waveguide sensor that significantly enhances particle-light interaction to improve sensitivity. The spiral waveguides are made of a chalcogenide glass material ($Ge_{23}Sb_7S_{70}$) which shows a broad transparency over IR range (1-10 μm). We demonstrate the sensing of N-methyl aniline-based aerosol particles with the device. We anticipate that the sensor will readily complement existing photonic resonator-based particle sizing and counting techniques to develop a unified framework for on-chip integrated photonic aerosol spectroscopy.

## 1. Introduction

The physicochemical characterization and monitoring of aerosols are crucial in many environmental and health-related settings. For instance, the physiochemical properties of pharmaceutical and diagnostic aerosols play a major role in determining the ease of dispersion, site of deposition, and route of clearance in the lungs [1,2]. Likewise, there are a number of unintended situations of potential aerosol exposure that result in risk to human health. Such cases include aerosols with living organisms such as fungi, bacteria or viruses. Specifically, bio-aerosols such as current pandemic scenario of COVID-19, SARC and other airborne viruses need tight environmental monitoring to ensure safe living atmosphere [3]. These applications necessitate ultra-sensitive methods to quantify aerosols for bacteriology and virology applications [4].

Various methods and instruments are available today to perform aerosol measurements ranging from filter based sample collection for offline laboratory analysis to sophisticated tools performing in-situ chemical analysis of aerosol particles [5]. Conventional methods that are employed for offline laboratory analysis of aerosols are based on Raman spectroscopy, FTIR spectroscopy, and fluorescence microscopy [6,7]. However, a significant challenge of these prevailing methods for chemical characterization is that they are either too bulky for field testing or have poor sensitivity when miniaturized to a hand-held form factor.

Recently, the development of ultra-sensitive miniaturized aerosol spectrometers has dramatically shaped industrial and academic research [8], with atmospheric and pharmaceutical industries investigating methods to characterize aerosols in real time [9]. Earlier this year, NASA supported an open challenge to develop an aerosol spectrometer for space shuttle applications that could probe the physicochemical properties of aerosol particles [10].

Prior efforts and research substantiate the need to change the system level design of the current sensing modalities where particle measurements are done in free space, because that requires sophisticated aerosol sampling and transport techniques. Also, the device needs to be designed to allow for miniaturization without compromising sensitivity. With the development of microphotonic fabrication technology, there is a possibility to shift to an on-chip photonic paradigm that could allow for ultra-sensitive spectroscopy of aerosols [11]. Near IR on-chip spectroscopic devices already exist for a variety of applications [11-15]. However, there is a need to develop similar photonic devices in the Mid IR range as most of the organic and inorganic aerosols show their chemical absorption in that range [12-18]. Over the past decade, several research groups have shown various MIR photonic platforms that are based on a variety of photonic materials and structures [19-27]. Miller et. al have shown Si based platform that could perform broad band spectroscopy in MIR range of 3-6 μm [21]. Singh et al. have used low stress SiN based platform for waveguide based MIR photonic sensing [25]. Al and Chen integrated crystalline silicon based micro ring resonators on mid-infrared compatible substrates for operation in the mid-infrared [22]. Sieger et al. designed the GaAs/AlGaAs thin film waveguides for MIR photonics [20]. Vasiliev et al. have demonstrated photo thermal based ultrasensitive MIR silicon on insulator photonic platform [23]. On these lines, Wang et al. summarized the recent work on III–V-on-silicon photonic integrated circuits based on the SOI platform in 2–4 μm wavelength range [26].

We propose an on-chip mid IR photonic aerosol spectrometer device that can identify the chemical attributes of the aerosol particles. The photonic sensor is developed using a chalcogenide glass material called $Ge_{23}Sb_7S_{70}$ that offers broad optical transparency over a wavelength range of 1μm to 10μm [11-15]. Other advantages of chalcogenide glasses are low processing temperatures and extensive alloying capabilities that can enable transparency beyond 10 μm wavelength for molecular fingerprinting [12-15]. Our sensor device (schematic shown in Fig. 1), consists of photonic waveguides forming a spiral structure to enhance the optical interaction of the particles with evanescent light. Such a design has a multitude of advantages.

First, it is a broadband device that allows us to sweep over a wide range of wavelengths. Second, the spiral structure in the device preserves the small footprint of the device without compromising the sensitivity of the measurements.

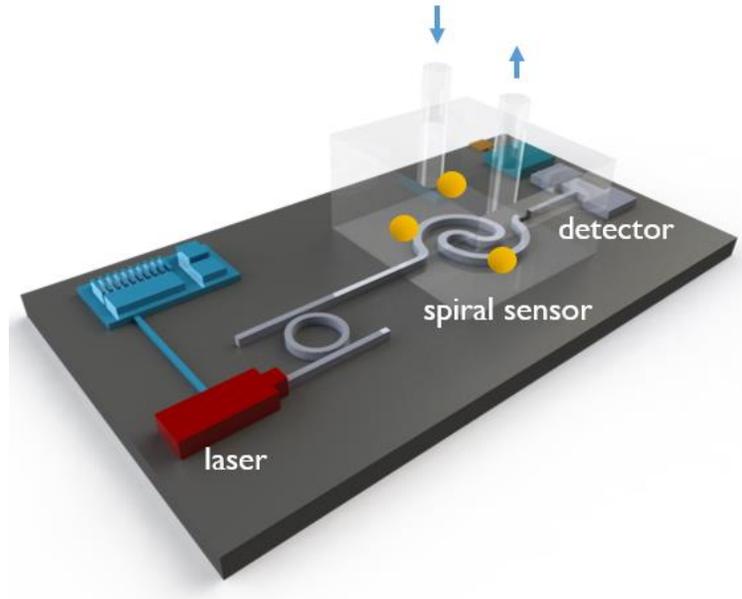

*Fig. 1: Schematic of our proposed on-chip MIR photonic aerosol spectrometer module.*

## 2. Device Design and Fabrication

We design a dual mode spiral waveguide structure for the MIR photonic sensor. Two aspects of the spiral waveguide needed to be optimized. First, the width and thickness of the waveguides, and second, the path length of the spiral to maximize the aerosol-light interaction with reduced waveguide losses. Here, we begin with making the cross-section of the spiral waveguide. We intend to support both polarization modes to enhance the light particle interaction. Hence, we design the waveguides to support the first two modes i.e., TE and TM. MODE [Lumerical Incorporated, Canada] simulations are performed to calculate the effective refractive index of the optical modes supported by the waveguide structure. The effective refractive index determining the mode confining capabilities of the waveguide, is analyzed for different width and thickness of the waveguide. The width and thickness of the waveguide are varied from 0.5 μm to 2 μm. Fig. 2a-b show the $n_{eff}$ map for the first two modes. The waveguide consists of three different layers, SiO2 as substrate ($n_s$=1.41), ChG glass material ($n_c$) as the waveguide core and air ($n_a$=1) as top cladding exposed to aerosols for sensing. The refractive index of the guiding layer must be higher the refractive indices of the substrate and air i.e. $n_c > n_s$, $n_c > n_a$. The condition requires $n_{eff}$ to be more than 1.41. Accordingly, Fig. 2a and Fig. 2b show the waveguide dimensions that could confine the modes partially and completely (distinguished by the dashed lines). Further, the waveguides are designed not to support any modes greater than 2. Fig. 2e shows $n_{eff}$ for first six modes in the waveguide of different width and 1.2 μm thickness. The left-top region (marked as mode cut region represents the waveguide width that supports only first two optical modes. Based on the analysis, we choose 1.5 μm wide and 1.2 μm thick waveguide (marked with asterisk in Fig. 2a and 2b) to fabricate the spiral structure.

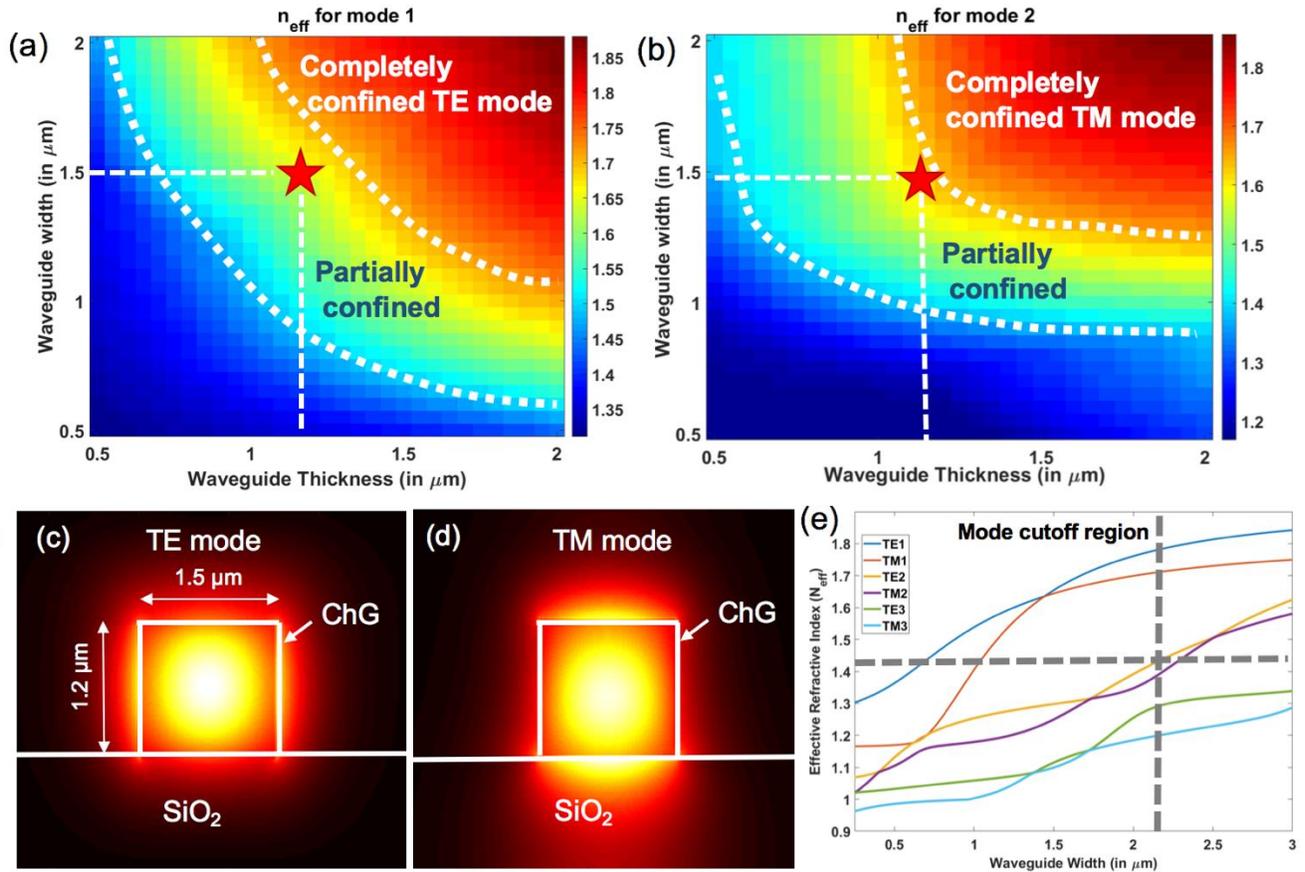

*Fig. 2: Design of the ChG spiral waveguides to support the first two optical modes (TE and TM) of input light wavelength of 3.1 μm to 3.4 μm. The mode is partially confined to allow larger evanescent tail interacting with the aerosol particles in vicinity. (a)-(b) represent the $n_{eff}$ map of the ChG waveguides for different width and thickness to support the TE and TM modes. Based on it, regions of complete and partial confinement are marked for TE and TM modes respectively. (c) Cross section view of the first TE mode in the spiral waveguide (d) Cross section of the first TM mode in the spiral waveguide. (e) For chalcogenide thickness of 1.2 μm, effective refractive index of first 6 modes versus waveguide width are shown.*

Since chalcogenide material absorption loss is low, we determine that the measured waveguide loss of 7 dB/cm is mainly due to scattering loss from sidewall roughness. We manage the trade-off of requiring a longer length for higher sensitivity, and a shorter length for lower loss, by choosing a total length of the sensor module (including the straight waveguide before and after the spirals structure), to be about 1 cm.

We pattern and fabricate the spiral waveguide structure using a double layer liftoff process which uses PMGI (MicroChem) as the undercut layer and ZEP-520A (ZEON) electron beam resist as the top layer. The fabrication process is outlined in Fig. 3. A silicon wafer with 3 μm thermal oxide is cleaned using a standard piranha solution immediately before a 1.5 μm thick PMGI layer and 400 nm thick ZEP layer is spun on top of it. The double layer is exposed using an Elionix ELS-F125 electron beam lithography tool. After the development of the pattern, we deposit a 1.2 μm thick layer of $Ge_{23}Sb_7S_{70}$ (GSS) glass using a thermal evaporator. The resist double layer is then lifted off in N-methyl-2-pyrrolidone, leaving the spiral structures.

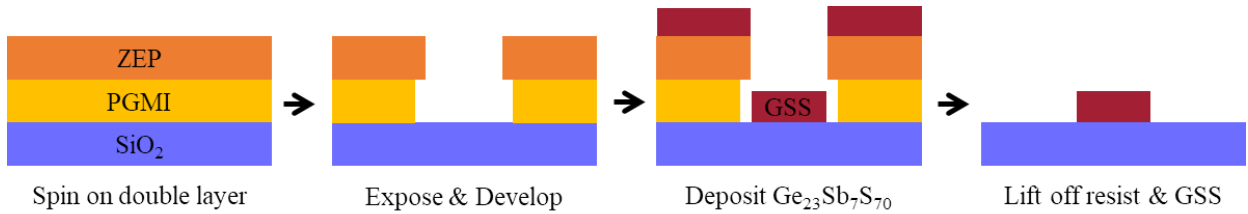

*Fig. 3. Schematic of fabrication process for spiral waveguides. A double layer made of PGMI and ZEP-520A is spun on the oxide wafer. After exposure to e-beam and developing the pattern, we deposit GSS (GeSbS chalcogenide material) and complete the fabrication by lifting off the resist and GSS.*

### 3. Device Characterization and Aerosol Sensing

The schematic of the experimental setup is shown in Fig. 4. The setup consists of two major components. First, the optical alignment setup. We couple light from a tunable MIR laser (Firefly from M-Squared) into the sensor module placed on a six-axis stage. Free space coupling is used because it allows easy handling of the aerosol delivery system to our sensor. A nitrogen cooled MIR camera (from IR Camera Inc.) detects the output light from the sensor module. A microscope on top assists in sensor alignment and monitors the aerosol interaction with the sensor.

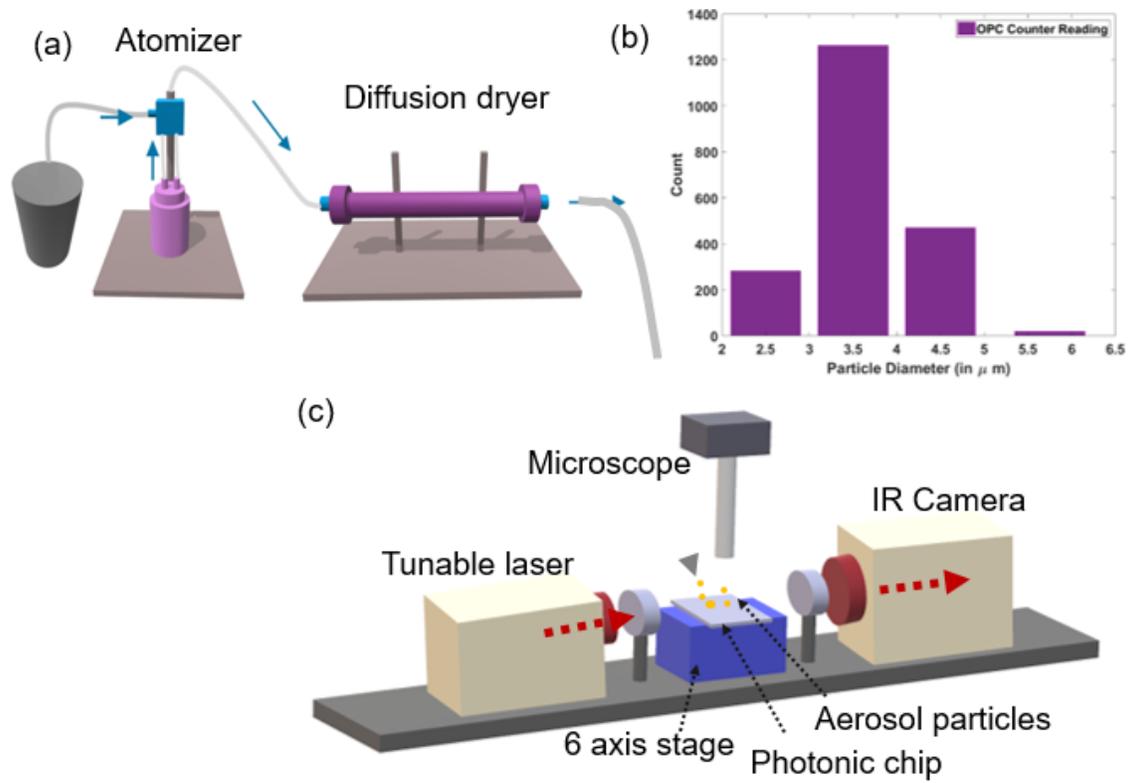

*Fig. 4: Schematic of the experimental setup. The setup consists of two major components (a & c ) i.e. (i) aerosol generation system that uses atomizer, diffusion dryer to generate aerosol mist. (ii) optical alignment system with tunable laser (Firefly from M-squared) and IR camera. The light is focused on to the photonic chip using the high NA lens. The photonic chip is placed on the 6 axis precision alignment stage. (b) Particle size and count obtained using commercial OPC particle counter from AlphaSense [28].*

The second major experimental component is the aerosol generating system. The aerosol generating mechanism uses a commercial compressed air based atomizer (TSI 3076) combined with a diffusion dryer (TSI 3062) to remove moisture from the atomized liquid. For our experiments we use N-methyl aniline liquid to generate atomized liquid droplets. The size and count of aerosol particles are represented in Fig. 4b. Fig. 5 displays a microscopic view of the sensor module with and without the aerosol exposure. It shows the aerosol particles falling on the surface and condensing around the waveguides.

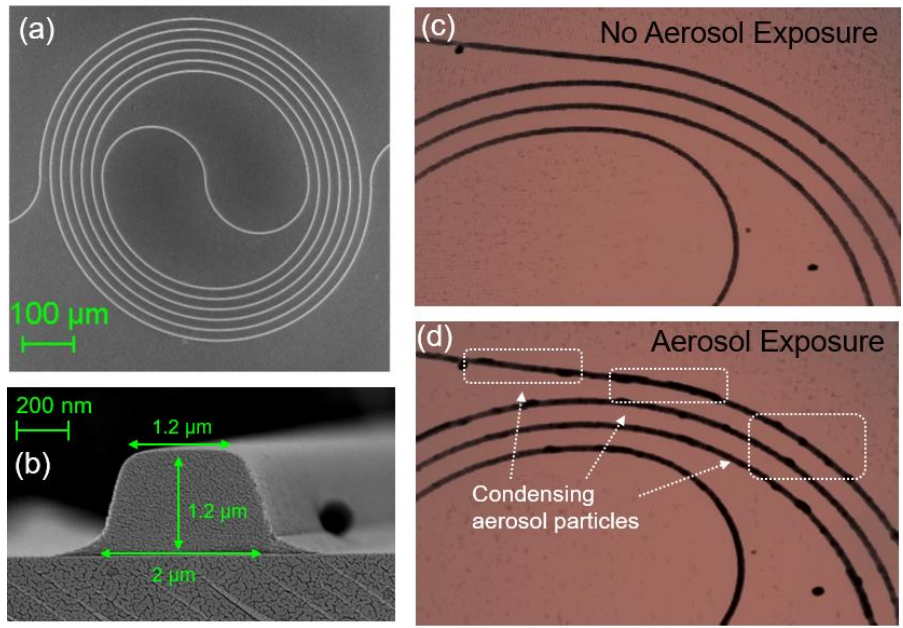

*Fig. 5: SEM and microscopic image of MIR chalcogenide spiral based aerosol sensor. (a) SEM image of the spiral top view (b) SEM image of the ChG waveguide cross sectional view (c,d) Microscopic view of the spiral sensor module with and without aerosol exposure. Condensing aerosol particles are observed on the waveguides as highlighted with the arrows.*

Once the broadband transmittance of the device over MIR is analyzed, we measure its response to the aerosol exposure. We perform a laser sweep manually from 3.25 µm to 3.35 µm of wavelength and measure the transmittance signal with the MIR camera. The obtained absorption spectrum of the aerosol particles is shown in Fig. 6. We normalize the absorption spectrum with the background noise to account for the air-borne contaminants. To assess the sensor performance, the spectrum is benchmarked with the data from NIST library (considered as a gold standard), and as exhibited in Fig. 6, our experimental absorption curve yields a good agreement in its shape with the gold standard. However, the sensor reading shows some fluctuations and artifacts in the absorption reading. We suspect the errors result from our manual sweep process during experimentation due to the dynamic nature of aerosol interaction with the sensor. To reduce such errors, we recommend an automatic sweep with the laser system which is beyond our current hardware capabilities at this time.

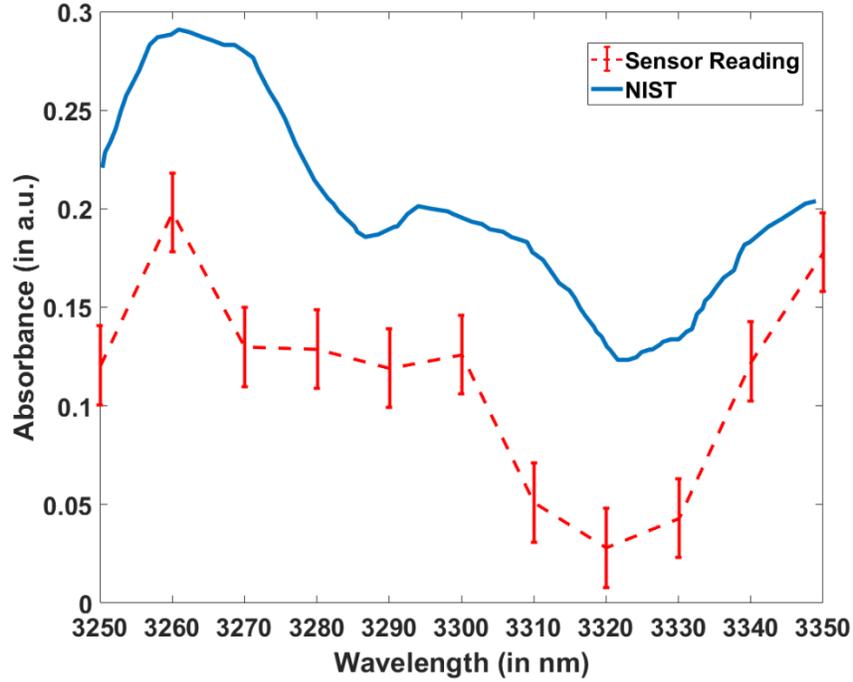

*Fig. 6: Absorption values of the N-methyl aniline based aerosol particles from the spiral based MIR photonic sensor. The absorption spectrum values are compared with the gold standard NIST absorbance values of the liquid N-methyl aniline. The comparison shows good agreement with the NIST spectrum with minor fluctuations. We attribute these fluctuations to the dynamic nature of aerosol while interacting with the photonic sensor module and the manual wavelength sweep performed during measurements.*

## CONCLUSION

In summary, we fabricated and characterized an on-chip MIR photonic platform for aerosol spectroscopy. Our sensor device performs chemical characterization of the aerosol particles using photonic spiral waveguide structures. The device design offers multiple advantages. First, the design enhances the particle light interaction with increased path length via a spiral structure. Second, the device is made of chalcogenide glass which is transparent over a wide range from near- to far infra-red. This allows us to use the sensor as a broadband device to perform on-chip infrared spectroscopy. Specifically, we illustrate the device functionality for N-methyl aniline-based aerosol particles. Further, we compare the absorption spectrum obtained from the device with one from the NIST library. The comparison yields good agreement with the sensor reading with minor fluctuations due to the dynamic nature of the aerosol environment.

The proposed device advances the on-chip photonic paradigm to perform MIR aerosol spectroscopy. It is particularly useful to progress and initiate miniaturized aerosol photonic sensors for a variety of applications in pharmacy, environmental sensing, and diagnostic applications. With an integrated photonic platform that includes a laser, sensing module, and detector together on-chip, we expect miniaturization and hence greater portability.

## ACKNOWLEDGMENTS

The authors acknowledge the infrastructure and support of Microsystems Technologies Laboratories (MTL), MIT to fabricate the photonic sensor.